\documentclass[conference]{IEEEtran}
\IEEEoverridecommandlockouts
\usepackage{cite}
\usepackage{amsmath,amssymb,amsfonts}
\usepackage{algorithmic}
\usepackage{graphicx}
\usepackage{booktabs}
\usepackage{multirow} 
\usepackage{comment}
\usepackage{textcomp}
\usepackage{xcolor}
\def\BibTeX{{\rm B\kern-.05em{\sc i\kern-.025em b}\kern-.08em
    T\kern-.1667em\lower.7ex\hbox{E}\kern-.125emX}}
\begin{document}

\title{DFALLM: Achieving Generalizable Multitask Deepfake Detection by Optimizing Audio LLM Components}

\author{
Yupei Li$^{*,1,4}$, Li Wang$^{*,2}$, 
Yuxiang Wang$^{2}$, Lei Wang$^{3}$, Rizhao Cai$^{3}$, Jie Shi$^{3}$, Bj\"orn W. Schuller$^{1,4}$, Zhizheng Wu$^{2}$ \\
\small{$^{*}$ Equal contribution.} \\
$^{1}$Imperial College London, UK \\
$^{2}$Chinese University HongKong, Shenzhen, China \\
$^{3}$Huawei, Singapore \\
$^{4}$Technical University Munich, Munich, German \\
}

\maketitle

\begin{abstract}
Audio deepfake detection has recently garnered public concern due to its implications for security and reliability. Traditional deep learning methods have been widely applied to this task but often lack generalisability when confronted with newly emerging spoofing techniques and more tasks such as spoof attribution recognition  rather than simple binary classification. In principle, Large Language Models (LLMs) are considered to possess the needed generalisation capabilities. However, previous research on Audio LLMs (ALLMs) indicates a generalization bottleneck in audio deepfake detection performance, even when sufficient data is available. Consequently, this study investigates the model architecture and examines the effects of the primary components of ALLMs, namely the audio encoder and the text-based LLM. Our experiments demonstrate that the careful selection and combination of audio encoders and text-based LLMs are crucial for unlocking the deepfake detection potential of ALLMs. We further propose an ALLM structure capable of generalizing deepfake detection abilities to out-of-domain spoofing tests and other deepfake tasks, such as spoof positioning and spoof attribution recognition. Our proposed model architecture achieves state-of-the-art (SOTA) performance across multiple datasets, including ASVSpoof2019, InTheWild, and Demopage, with accuracy reaching up to 95.76\% on average, and exhibits competitive capabilities in other deepfake detection tasks such as attribution, and localisation compared to SOTA audio understanding models. Data and codes are provided in supplementary materials.
\end{abstract}


\section{Introduction}
\label{sec:intro}


Audio deepfake detection has become a significant public concern regarding security and reliability. Traditional detection methods, often based on smaller audio models like WavLM \cite{Chen_2022}, achieve high accuracy on known datasets. However, they suffer from two critical limitations. First, they lack generalizability; their performance degrades sharply when confronted with newly emerging spoofing techniques or out-of-domain (OOD) data \cite{yi2023audio}. Second, they are single-task models, struggling to handle multiple tasks like spoof attribution or localization within a single model.

In principle, Audio Large Language Models (ALLMs) are an ideal solution to these challenges. Their large parameter count suggests greater generalization potential. Furthermore, their prompt-based nature makes them inherently suited for complex, multitask instructions (e.g., detection, attribution, and localization).

However, when directly applied to deepfake detection, existing ALLMs still exhibit a generalization bottleneck. We argue that this phenomenon is due to the information bottleneck in their audio encoders and corresponding textual LLMs \cite{gu2025allm4addunlockingcapabilitiesaudio}.

Currently, most of ALLMs utilise Whisper \cite{radford2022robustspeechrecognitionlargescale} as their audio encoders \cite{chu2023qwenaudioadvancinguniversalaudio, ding2025kimi, alex2025palprobingaudioencoders}. However, Whisper is typically trained with supervision on ASR tasks. This training objective compels the model to preserve linguistic content while simultaneously discarding non-linguistic, acoustic details deemed irrelevant to speech recognition. Critically, it is precisely this discarded acoustic information that has been shown to be essential for robust deepfake detection.

This analysis leads to our core hypothesis: the key bottleneck for ALLM generalization is an encoder-level problem, not an LLM-level one. We therefore posit that the key to unlocking performance is not the LLM, but the choice of the encoder. To verify this, we conduct a systematic investigation of the core ALLM components, explicitly comparing semantic-optimized encoders (like Whisper) against acoustically-aware encoders (like Wav2Vec2-BERT \cite{chung2021w2vbertcombiningcontrastivelearning}) which preserve more raw signal features through self-supervised masked modeling.

Based on these findings, we propose DFALLM, a framework optimized for generalizable, multitask deepfake detection. \textbf{Our main contributions are threefold}:
First, we demonstrate through systematic experiments that the Audio Encoder is the primary performance bottleneck for ALLMs in deepfake detection. We show that an acoustically-aware encoder (Wav2Vec2-BERT) significantly outperforms a semantic-optimized encoder (Whisper) in generalization. Second, we propose an optimized ALLM architecture (DFALLM) that achieves state-of-the-art (SOTA) detection performance across multiple benchmarks, including ASVSpoof2019, In The Wild, and Demopage, by optimizing component selection, frame rate, and training strategy. Third, we design a unified, prompt-based multitask framework. We show that an ALLM is superior for handling detection, attribution, and localization in a single model. Notably, DFALLM far exceeds smaller models on complex reasoning tasks (localization), validating the necessity of the LLM for comprehensive forensic analysis.

\begin{figure*}[!t]
    \centering
    \includegraphics[width=0.8\linewidth]{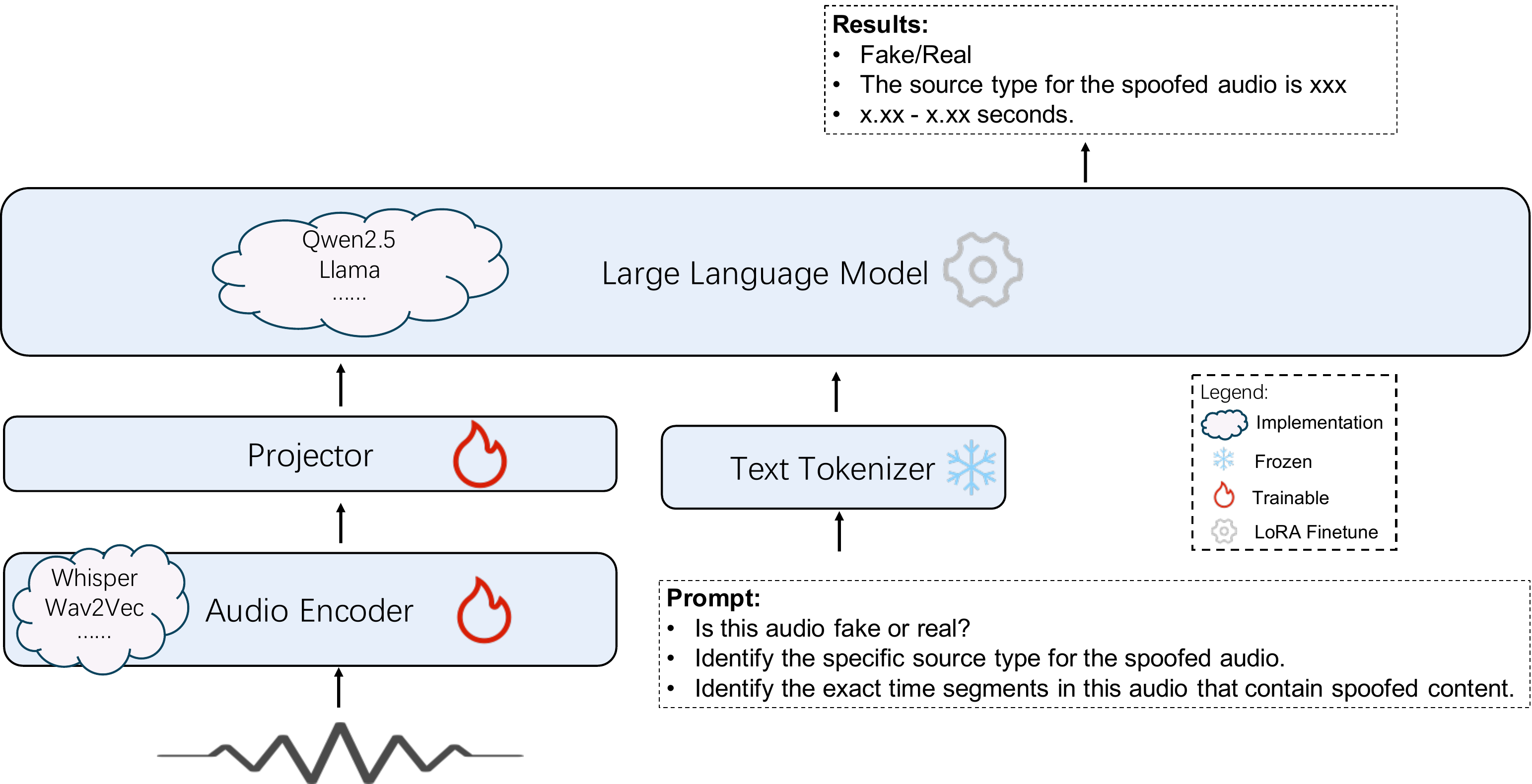}
    \caption{DFALLM: The proposed framework takes audio inputs with varying frame rates alongside corresponding textual prompts. The audio signals are first processed by an audio encoder, whose output representations are projected into the text embedding space. These projected embeddings are then combined with the representations obtained from the text tokenizer and jointly processed by a textual LLM to interpret the audio–text input pair and generate task-specific responses. In our setup, the audio encoder is fully trained, while the textual LLM is fine-tuned using the LoRA technique to efficiently adapt its parameters. Both components are modular and can be selected from existing model libraries to enable systematic investigation of their respective contributions.}
    \label{fig:pipeline}
\end{figure*}

\section{Related work}

Current deepfake detection research can be grouped into three subtasks: binary detection, attribution, and localization. Binary detection focuses on distinguishing bona fide from spoofed audio, with early work relying on deep learning models such as Whisper or spectrogram-based classifiers~\cite{yi2023audio}; AASIST~\cite{Jung2021AASIST} and post-training approaches~\cite{ge2025posttrainingdeepfakespeechdetection} further advanced performance. Attribution identifies which generative model produced a deepfake~\cite{Klein_2024}, leveraging similar techniques, including digital fingerprint analysis~\cite{Yan_2022} and large-scale benchmark studies~\cite{xie2025neuralcodecsourcetracing}. Localization aims to determine the precise manipulated segments~\cite{he2025manipulated}, with compact models proposed to capture fine-grained feature patterns~\cite{zhang2024mfms}. Despite these developments, prior work typically addresses the three subtasks in isolation, remains largely unimodal, and continues to face challenges in generalization.

With the rise of ALLMs such as AudioPaLM \cite{rubenstein2023audiopalmlargelanguagemodel}, joint speech–text reasoning has shown strong results in ASR and audio captioning, yet their use in forensic audio, particularly deepfake detection, remains limited~\cite{gu2025allm4addunlockingcapabilitiesaudio}. Prior work suggests that fine-tuning the audio encoder is key, rather than keeping it frozen, but evaluations have focused solely on ASVspoof 2019, with no OOD testing, leaving generalizability unclear.

ALLM has the potential of multitask learning and generalization with large number of parameters. Previous ALLMs have demonstrated strong performance across a wide range of audio understanding tasks, such as speech recognition, audio captioning, and spoken question answering \cite{das2024speechverse, yang2024uniaudio}. A key factor to their success is their integration of textual LLMs, such as Qwen 2 \cite{team2024qwen2} and LLaMA \cite{touvron2023llama}, as foundational components. These LLMs bring knowledge and reasoning and generalization abilities. ALLMs can hence effectively bridge acoustic inputs with high-level language semantics. However, multitask learning in deepfake analysis that combines detection, attribution, and localization has been little explored.

Therefore, the primary aim of this paper is to investigate the capability of ALLMs in deepfake detection. In addition, we explore an important open question concerning the scaling behavior of large models~\cite{held2025relative}, which suggests that model capacity and data size jointly determine overall performance in large language models. Our work empirically examines these factors to identify the optimal model scale, as their influence on the deepfake detection ability of ALLMs remains insufficiently understood.

\section{Methodology}

This section details the DFALLM framework, its multitask prompting strategy, and the systematic investigation strategy used to analyze its core components.

\subsection{DFALLM Architecture}

The architecture of our proposed framework, DFALLM, is illustrated in Figure~\ref{fig:pipeline}. DFALLM is a speech language model composed of three core components: an audio encoder, a text tokenizer, and a textual LLM. The model accepts both audio and text as input and generates text as output.

The data flow utilizes these components in parallel. The audio encoder processes the raw audio signal to extract acoustic representations. Simultaneously, the text tokenizer converts the textual prompt into token embeddings. Finally, the textual LLM jointly processes the audio representations (mapped via a projection module) and the text embeddings to generate a task-specific response.

\subsection{Investigation Strategy}
Our investigation strategy employs a sequential, two-stage process to validate the model's capabilities.

Stage 1: Component Analysis for Generalization. The first stage focuses on the primary deepfake detection task to identify the optimal components for generalization. We test our core hypothesis by systematically comparing configurations of the two main components: \textbf{Audio Encoders}, contrasting semantic-optimized with acoustically aware designs, and \textbf{Textual LLMs}, evaluating the influence of different model families and scales on detection performance.

Stage 2: Multitask Validation. In the second stage, we take the single, optimal component configuration identified from Stage 1. We then apply this unified model to the full suite of forensic tasks (detection, attribution, and localization) to evaluate its effectiveness as a generalized, multitask framework.

\subsection{Multitask Prompting Strategy}
A key capability of DFALLM is performing multiple audio deepfake tasks within a single, unified model. This is achieved by guiding the LLM with task-specific prompts. This approach allows the model to generalize its capabilities across different audio deepfake-related tasks.

We designed distinct prompts for three primary tasks:
For Detection (Binary): \emph{Is this audio fake or real?} For Attribution (N-Class): \emph{Identify the specific source type or the spoofed audio.} For Localization: \emph{Identify the exact time segments in this audio that contain spoofed content.}

\section{Experiments}

\subsection{Datasets}
\textbf{Training Corpus:} We built a combined training dataset by sampling from multiple public deepfake audio datasets, including ASVspoof 2019 \cite{wang2020asvspoof}, SpoofCeleb \cite{jung2025spoofcelebspeechdeepfakedetection}, MLAADv6 \cite{müller2025mlaadmultilanguageaudioantispoofing}, ReplayDF \cite{muller2025replaydf}, DFADD \cite{du2024dfadddiffusionflowmatchingbased}, AISHELL3 \cite{shi2021aishell3multispeakermandarintts}, ADD2023 \cite{yi2023add2023secondaudio}, GigaSpeech \cite{Chen_2021}, CNCeleb \cite{fan2019cncelebchallengingchinesespeaker}, PartialSpoof \cite{Zhang_2023}.

\textbf{Evaluation Sets:} Evaluation Sets: We used distinct evaluation sets for each task.
  For Detection: To assess both in-domain and out-of-domain (OOD) performance, we employed three datasets. ASVSpoof 2019 LA (71,237 samples) serves as the in-domain set . In-the-Wild (ITW) \cite{muller2022does} (31,780 samples) and Demopage (2,182 aggregated samples) are used as OOD test sets. These are provided in supplementary materials, with Demopage meaning the collection from samples shown on the demo page of various TTS models.
 
  For Attribution: The test set was a combination of ASVspoof2019 LA, SpoofCeleb, DFADD, ReplayDF, and ADD2023, totaling 3,836 data samples.
   
    For Localization: We used the PartialSpoof test set, sampling 1,000 instances with annotated spoofed periods.

\subsection{Model Configurations}

For the investigation of ALLM components, we selected specific open-source pretrained models. Two representative \textbf{audio encoders} were considered. \emph{Whisper} (small, medium, and large-v3 variants) is trained with ASR supervision. \emph{Wav2Vec2-BERT} is a model for standalone deepfake detection, outperforming Whisper on that task; however, its integration within ALLMs has not been previously explored. Each encoder was augmented with a simple linear classification head and used as a baseline in the ALLM experiments.

We also evaluated several \textbf{textual LLMs}, including members of the Qwen family (Qwen2.5-0.5B, Qwen2.5-1.5B, Qwen2.5-7B, and Qwen3-0.6B) as well as Llama-1.3B, to examine the impact of model family and scale on detection performance. Finally, to assess the influence of temporal resolution, audio \textbf{frame rates} of 12.5 Hz (native to Whisper) and 50 Hz (native to Wav2Vec2) were tested.

\subsection{Training and Implementation Details}

A projection module is used to map the audio representations from the audio encoder into the text embedding space. We use a Linear layer as the projection module. For our training strategy, we fully trained the audio encoder and projector to adapt it to the acoustic features of the task. The LLM was then efficiently fine-tuned using Low-Rank Adaptation (LoRA) \cite{hu2021loralowrankadaptationlarge}.

\begin{table}[htbp]
\centering
\small
\caption{Training hyperparameters}
\resizebox{0.40\textwidth}{!}{
\begin{tabular}{lp{0.27\textwidth}}
\toprule
\textbf{Parameter} & \textbf{Value / Setting} \\
\midrule
lora\_rank & 16 \\
lora\_alpha & 32 \\
lora\_target\_modules & 
projection of q,k,v,o,gate,up,down \\
max\_tokens & 512 \\
max\_eval\_samples & 4000 \\
\midrule
train\_epochs & 4 \\
learning\_rate & 1e-5 \\
lr\_scheduler\_type & cosine \\
seed / data\_seed & 42 / 42 \\
model\_max\_length & 1024 \\
\bottomrule
\end{tabular}}
\label{tab:hyperparams}
\end{table}

All models were trained for 4 epochs with a cosine learning rate scheduler and a learning rate of 1e-5. For LoRA, rank is set to 16. The key hyperparameters are summarized in Table~\ref{tab:hyperparams}.

We experimented with multiple audio encoders, including Wav2Vec2-Bert \cite{chung2021w2vbertcombiningcontrastivelearning} and Whisper \cite{radford2022robustspeechrecognitionlargescale}, to extract robust acoustic representations from speech signals. For the textual component, we explored different LLMs, including Qwen2.5 \cite{team2024qwen2}, Qwen3 \cite{yang2025qwen3technicalreport} and LLaMA \cite{touvron2023llamaopenefficientfoundation}, to process both the projected audio representations and the accompanying textual prompts. 

We selected these models because the chosen audio encoders are representative of current architectures, each emphasizing different aspects of audio understanding. Whisper prioritizes semantic comprehension, as it is trained primarily on large-scale ASR corpora, whereas Wav2Vec2-BERT, derived from Wav2Vec2 \cite{baevski2020wav2vec20frameworkselfsupervised}, leverages self-supervised masked language modeling, thereby preserving more of the raw acoustic and physical characteristics of the speech signal. Comparing these two encoders allows us to identify potential bottlenecks in audio representation that may hinder the generalizability of ALLMs in deepfake detection. For the textual LLMs, we selected two models from the Qwen family and one from the LLaMA family to maintain diversity in the model pool while enabling controlled comparisons across architectures with differing linguistic priors and adaptation strategies.

To enhance task alignment, we designed task-specific prompts that guide the LLM during the detection process, thereby improving the model’s generalization across various deepfake-related tasks, including detection, attribution, and localization. The resulting system demonstrates consistent and robust performance across multiple deepfake detection tasks. The prompts are simple and shown in Figure \ref{fig:pipeline}.


After establishing the pipeline to investigate the influence of different audio encoders and textual LLMs, we aim to answer \textbf{what is the optimised configuration of encoder and textual LLM selection}, and \textbf{which component has more influence to make ALLMs generalizable for OOD and perform multitasks}. Audio encoders influence the representation of the audio, while textual LLMs influence how model processes the information. We therefore evaluated the Whisper-small and Wav2Vec2-BERT audio encoders, which have comparable parameter sizes, with the same textual LLM. We then examined Wav2Vec2-BERT paired with different textual LLMs.

\section{Results and Analysis}

\subsection{Main findings}
We report average accuracy (Acc)
, together with in-domain (ID) test accuracy (ASVSpoof2019) and average accuracy on OOD test accuracy (Demopage and ITW) for binary deepfake detection in Table~\ref{tab:audio_llm_accuracy}. For other experiments, overall average accuracy is shown as an indicator of general performance.
\begin{table}[ht]
\centering
\caption{Performance comparison of different Audio Encoder and LLM configurations.}
\label{tab:audio_llm_accuracy}
\resizebox{0.48\textwidth}{!}{

\begin{tabular}{l l c c c}
\toprule
\textbf{Audio Encoder} & \textbf{LLM} & \textbf{Average Acc (\%)} & \textbf{ID Acc (\%)} & \textbf{OOD Acc (\%)} \\
\midrule
\multicolumn{5}{c}{\textit{Audio Encoder Only (Baseline)}} \\
\midrule
Whisper (small) & N/A & 84.05 & 98.45 & 76.85 \\
Wav2Vec2-BERT & N/A & 94.89 & 99.15& 92.76\\
\midrule
\multicolumn{5}{c}{\textit{Audio Encoder + Qwen2.5-0.5B}} \\
\midrule
Whisper (small) & Qwen2.5-0.5B & 84.89 & 95.35& 79.66 \\
Whisper (medium)     & Qwen2.5-0.5B &  84.79&97.75& 78.31\\
Whisper (large-v3)   & Qwen2.5-0.5B &  90.87&96.06& 88.28\\
Wav2Vec2-BERT & Qwen2.5-0.5B & \textbf{95.76} & \textbf{99.15}& \textbf{94.07}\\
\midrule
\multicolumn{5}{c}{\textit{Wav2Vec2-BERT + LLM}} \\
\midrule
Wav2Vec2-BERT & Qwen2.5-0.5B & \textbf{95.76}&\textbf{96.20}& 94.07\\
Wav2Vec2-BERT & Qwen3-0.6B & 94.86 &95.77& \textbf{94.47}\\
Wav2Vec2-BERT & Qwen2.5-1.5B & 93.79&95.92&92.73 \\
Wav2Vec2-BERT & Qwen2.5-7B & 93.94&96.20& 92.81\\
Wav2Vec2-BERT & Llama-1.3B & 85.67&94.08& 81.46\\
\bottomrule
\end{tabular}}
\end{table}

\textbf{It can be observed that the choice of audio encoder serves as the decisive factor in overall model performance.} When evaluated independently, Wav2Vec2-BERT substantially outperforms Whisper-small, achieving accuracies of 94.89\% and 84.05\%, respectively. This indicates that the quality of audio feature extraction plays a dominant role in downstream classification accuracy, and that Wav2Vec2-BERT is better suited for the deepfake detection task. When integrated with a textual LLM, the Wav2Vec2-BERT–based system continues to outperform the Whisper-based one by around 10\%, showing the importance of the audio encoder selection.

Moreover, when the audio encoder is held the same, different LLMs yield varying results, demonstrating that the choice of textual model also influences performance, with Qwen2.5-0.5B emerging as the most effective configuration. The inclusion of a textual LLM provides additional performance gains, although the magnitude of improvement is relatively modest compared to only using an audio encoder. Overall, the results suggest that variations in the audio encoder exert a greater influence on model performance than modifications to the LLM. Additionally, more advanced LLMs consistently outperform smaller, traditional models.

Another important conclusion concerns the relative impact of model scale across the audio encoder and the textual LLM. \textbf{For the audio encoder, bigger is clearly better.} Analysis of Audio Encoder + Qwen2.5-0.5B configurations shows performance rises substantially with encoder size: Whisper large-v3 achieves 90.87\%, outperforming small (84.89\%) and medium (84.79\%) variants. This indicates high-capacity encoders are crucial for capturing subtle spectral and temporal artifacts of deepfake audio. In contrast, \textbf{for the textual LLM, lightweight is sufficient.} Wav2Vec2-BERT + LLM results show the smallest model, Qwen2.5-0.5B, performs best (95.76\%), exceeding larger models like 1.5B (93.79\%) and 7B (93.94\%). Scaling up the LLM provides little benefit and may even degrade performance. These findings suggest effective deepfake detection relies on the encoder’s high-capacity acoustic modeling, while the textual LLM mainly contributes semantic reasoning and can be small. The LLM remains essential, enhancing performance even at a modest scale.

In conclusion, the optimal configuration is the Wav2Vec2-BERT + Qwen2.5-0.5B combination, which achieves the highest overall accuracy among all tested settings. This pairing effectively balances the representational power of the audio encoder with the contextual reasoning capabilities of a lightweight LLM, resulting in superior performance.

Although the aforementioned ALLM configurations demonstrate some generalizability on out-of-distribution datasets, it remains necessary to evaluate their performance across multiple tasks, including attribution and localization. Multitask evaluation provides a rigorous test of whether the models truly understand and analyze the audio signals that indicate a deepfake, rather than relying on spurious correlations or chance~\cite{yi2023audio}. A more comprehensive assessment is therefore required to determine if the models can accurately identify both the presence of a deepfake and its underlying characteristics. Our objective is to empirically assess whether ALLMs achieve superior performance not only on individual tasks but also in a multitask setting encompassing attribution and localization.

We fine-tuned the best-performing configuration on a multitask dataset covering detection, attribution (19 classes), and localization. For comparison, we also trained the audio encoder independently, using two classifiers for detection and attribution, and a regressor for predicting the start and end times of manipulated segments. Results are in Table \ref{tab:detection-attribution-localization}, with accuracy reported for detection and attribution, and localization evaluated via Intersection over Union (IoU) between predicted and ground-truth temporal regions. Training parameters matched the single-task setup; only the dataset and prompts differed.

\begin{table}[ht]
\centering
\caption{Performance comparison of ALLMs and small models on detection, attribution, and localization tasks.}
\resizebox{0.48\textwidth}{!}{
\begin{tabular}{lccc}
\toprule
\textbf{Model} & \textbf{Detection (\%)} & \textbf{Attribution (\%)} & \textbf{Localization (IoU)} \\
\midrule
Wav2Vec2-BERT+Qwen2.5-0.5B & \textbf{98.67} & 86.98 & \textbf{74.00} \\
Wav2Vec2-BERT               & 98.02 & 92.69 & 53.84 \\
\bottomrule
\end{tabular}}
\label{tab:detection-attribution-localization}
\end{table}

Our experimental results demonstrate that \textbf{ALLMs, under the proposed training configuration, achieve better generalization across deepfake detection multitasks including detection, attribution, and localization.} The integration of language understanding capabilities significantly enhances the model’s global perception of speech forgery characteristics, enabling a deeper and more contextualized interpretation of audio features. Although the attribution task makes the ALLM perform slightly worse than smaller models, this is due to the inherent next-token-prediction behavior of LLMs. Importantly, this does not weaken the claim of the model’s audio understanding or generalization ability, as demonstrated by its overall multitask performance. Besides, the results indicate that smaller models without textual LLM components struggle to generalize effectively to multitask settings, thereby validating the necessity of ALLMs for robust deepfake detection.

\subsection{Ablation study}
After presenting the main results, we further examine two critical factors that may influence the effectiveness of audio–language models for deepfake detection which are also essential in ALLMs configurations:the frame rate used during audio feature extraction and the amount of training data. 

First, audio representations are influenced not only by the choice of the audio encoder but also by the characteristics of the original audio input. Previous explainable AI studies have shown that certain distinguishable deepfake features are concentrated within specific frequency ranges of the audio \cite{maltby2024frequency, feng2024audios, li2024detectingmachinegeneratedmusicexplainability}. Moreover, different audio encoders operate at varying frame rates, which affect the effective frequency resolution of the input; for example, Whisper uses a frame rate of 12.5\,Hz, whereas Wav2Vec2 operates at 50\,Hz. Consequently, we also investigate the optimal framerate as a component of our overall configuration, with results shown in Table \ref{tab:frame-rate-results}. 

\begin{table}[h]
\centering
\caption{Effect of the frame rate on deepfake detection performance across datasets. The table reports accuracy for average accuracy.}
\begin{tabular}{l l c c}
\hline
\textbf{Audio Encoder} & \textbf{LLM} & \textbf{Frame Rate} & \textbf{Accuracy (\%)} \\
\hline
\multirow{2}{*}{Whisper (large-v3)} & \multirow{2}{*}{Qwen2.5-0.5B} & 12.5Hz & 86.94 \\
                                   &                                  & 50Hz    & \textbf{90.87}\\
\hline
\multirow{2}{*}{Wav2Vec2-BERT}     & \multirow{2}{*}{Qwen2.5-0.5B}   & 12.5Hz & 95.70 \\
                                   &                                  & 50Hz    & \textbf{95.76} \\
\hline
\end{tabular}
\label{tab:frame-rate-results}

\end{table}

Overall, higher frame rates yield richer audio representations, which in turn enhance detection accuracy. We hypothesize that the temporal granularity of the audio representation plays a crucial role. Whisper, optimized for semantic compression, operates at a lower frame rate (12.5 Hz), effectively smoothing out high-frequency spectral artifacts and transient glitches common in deepfakes. 
In contrast, Wav2Vec2-BERT maintains a higher temporal resolution (50 Hz). While the performance gap is primarily driven by the pre-training objectives (acoustic vs. semantic), the finer temporal granularity of 50 Hz serves as a necessary condition for capturing subtle, short-duration manipulation traces. Our results in Table \ref{tab:frame-rate-results} confirm that models with higher native temporal resolution (Wav2Vec2-BERT) consistently outperform those with compressed temporal latents (Whisper).

\begin{figure}[h]
    \centering
    \includegraphics[width=0.95\linewidth]{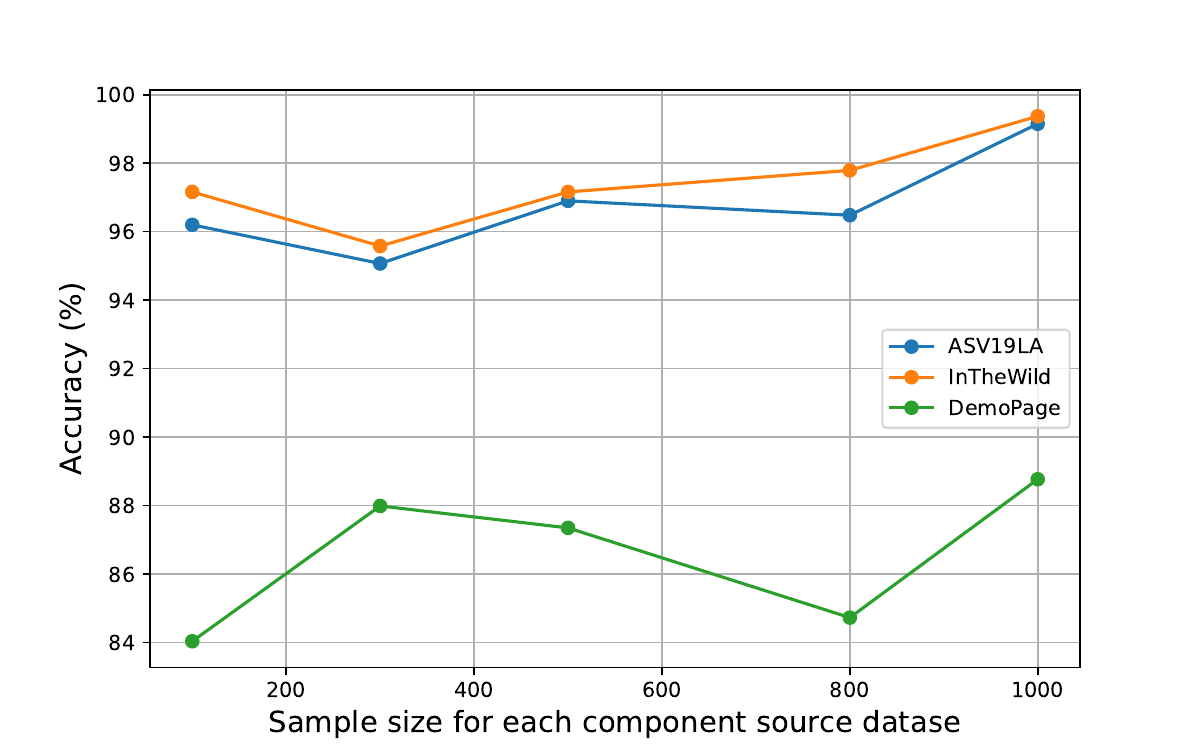}
    \caption{Performance on Qwen2.5-0.5B with Wav2Vec-Bert across datasets with different sample size for each component source dataset}
    \label{fig:datasize}
\end{figure}

Moreover, although a number of parameters is required to effectively perform the deepfake detection task, we conducted experiments to evaluate whether a small amount of training data is sufficient to achieve comparable performance. To this end, we created training subsets of varying sizes by sampling from 100, to 1000., instances from each component training dataset. This allows us to analyze the effect of training data volume, given that the full dataset contains over 170k samples, with results shown in Figure \ref{fig:datasize}.

These results reflect a behavior that fits for LLM scaling law \cite{kaplan2020scalinglawsneurallanguage}: performance improves steadily as more training data is provided, even though the underlying LLM remains small. It also indicates that achieving robust performance with LLMs is not straightforward, as the task requires sufficient data for the model to fully understand the audio and accurately identify potential deepfaked segments. While smaller models can perform reasonably well on relatively constrained datasets, larger models are able to realize their full potential when provided with more extensive training data. This may be because deepfake detection demands a genuine understanding of the audio content rather than superficial signal-based guesses. These findings further suggest that ALLMs are capable of learning intrinsic audio features that support effective detection across varying conditions.



\section{Conclusion}

In this work, we analyzed the roles of the audio encoder and textual LLM in shaping ALLMs’ generalization across datasets and multitask deepfake detection, attribution, and localization. Our results show that Wav2Vec2-BERT combined with Qwen2.5-0.5B achieves the best overall detection performance. We further assessed the impact of frame rate, model size, and training data volume. Future work will extend this configuration to broader audio understanding tasks to probe ALLMs’ generalization in complex multimodal scenarios.

\bibliographystyle{IEEEbib}
\bibliography{icme2026references}

\end{document}